\documentclass[pdftex,twocolumn,epjc3]{svjour3}          
\smartqed  
\RequirePackage{graphicx}

\usepackage{bm}
\usepackage{amsbsy}
\usepackage{amssymb}
\usepackage{physics}

\RequirePackage{latexsym}
\RequirePackage[numbers,sort&compress]{natbib}
\RequirePackage[colorlinks,citecolor=blue,urlcolor=blue,linkcolor=blue]{hyperref}

\def\ee{\varepsilon}

\newcommand{\I}{\mathrm{i}} 
\newcommand{\D}{\mathrm{d}} 
\newcommand{\E}{\mathrm{e}} 

\def\sinh{\mathrm{sinh}}

\def\cotanh{\mathrm{cotanh}}

\def\diag{\mathrm{diag}}
\def\Re{\mathrm{Re}}
\def\arg{\mathrm{arg}}

\journalname{Eur. Phys. J. C}
\begin{document}

\title{Position-space representation of charged particles' propagators in a constant magnetic field as an expansion over Landau levels}


\titlerunning{Position-space representation of propagators}        

\author{Iablokov S.N.\thanksref{e1,addr1}
        \and
        Kuznetsov A.V.\thanksref{addr1} 
}

\thankstext{e1}{e-mail: physics@iablokov.ru}


\institute{P.G. Demidov Yaroslavl State University, Yaroslavl, Russia  \label{addr1}}

\date{Received: date / Accepted: date}

\maketitle

\begin{abstract}
We have obtained propagators in the position space as an expansion over Landau levels for the charged scalar particle, fermion, and massive vector boson in a constant external magnetic field. The summation terms in the resulting expressions consisted of two factors, one being rotationally invariant in the 2--dimensional Euclidean space perpendicular to the direction of the field, and the other being Lorentz-invariant in the 1+1--dimensional space-time. The obtained representations are unique in the sense that they allow for the simultaneous study of the propagator from both space-time and energetic perspectives which are implicitly connected. These results contribute to the development of position-space techniques in QFT and are expected to be of use in the calculations of loop diagrams.

\end{abstract}

\section{Introduction}

In the vast majority of QFT problems, the momentum-space paradigm is adopted due to the simplicity of respective calculations and the ability to apply a variety of regularization methods. However, for some problems (e.g., sunrise-type diagrams) the position-space techniques allow for a much simpler \cite{Groote:1999cx,Groote:2004qq,Groote:2005ay,Groote:2012pa} evaluation of integrals. These integrals (which are called Bessel moments \cite{bailey2008elliptic}) consist of products of different types of Bessel functions. There exist a plethora of analytic expressions for such integrals containing a product of two, three, four, and even more Bessel functions as integrand \cite{GR,PBM}. For integrals with unknown closed analytic forms, numerical analysis remains a feasible choice due to the underlying symmetries allowing to reduce the dimension of integration space to one.

The analysis of divergences (which arise in loop processes) is also possible in the position space. In several papers \cite{Bollini:1995pp,Plastino:2017kgl} there was discussed the dimensional regularization procedure for the position-space expressions. It consisted in the control of the degree $\nu$ of the respective Bessel functions which depend on $\nu$ analytically. Another possibility for a correct treatment of infinities, namely a cutoff in the position space, was also reported \cite{Groote:2018rpb}.

Finally, the position-space calculations are indispensable when it comes to studying processes with space-dependent external fields and initial-state wavepackets of arbitrary shapes, e.g., when considering finite-spacetime problems, such as those related to neutrino oscillations \cite{akhmedov2011neutrino,falkowski2020consistent}.

All of this demonstrates the growing significance of position-space methods in QFT and serves as a motivation for obtaining the most important ingredients for these calculations, i.e., the position-space expressions of particles' propagators. Such formulas for free particles were well known for a long time \cite{Bogoliubov_QFT,Greiner_QED}. Other valuable examples of position-space propagators were also considered in the literature, namely, the propagator of the Dirac equation in an external plane wave \cite{DiPiazza:2018ofz} and propagators of charged particles in a constant magnetic field. In the latter case, the list of published representations includes (i) the proper-time \cite{Fock:1937} representation in the position space \cite{Schwinger:1951,Itzykson_1980,gavrilov1996proper,gavrilov1998qed,gavrilov2004green}, (ii) the proper-time representation in the momentum space \cite{KM_Book_2013,Erdas_1990,Erdas_2000} and (iii) the Landau-levels representation in the momentum space \cite{Chodos_1990,Chyi_2000,Nikishov,KM_Book_2013}. To the best of our knowledge, no Landau-levels position-space formulas were reported, except for our previous work where we presented the position-space representation of the charged scalar particle propagator in a constant magnetic field expanded as a sum over the Landau levels \cite{Iablokov:2020ypp}.

Landau-levels expansion of a charged particle propagator allows for a straightforward interpretation. Each term in the series has a factor corresponding to a 1+1--dimensional particle with a modified mass that depends on the Landau level (see Chapter 2). Under certain conditions, a truncation of the series is possible, allowing to achieve a confident approximation with a compact closed-form expression. However, there exist several precedents when misunderstanding of such conditions led to wrong conclusions. For instance, a calculation of the neutrino self-energy operator in a magnetic field was performed in Refs.~\cite{Elizalde:2002,Elizalde:2004} by analyzing the one-loop diagram $\nu \to e^-\, W^+ \to \nu$. The authors restricted themselves to the contribution to the electron propagator from the ground Landau level. As it was shown in Ref.~\cite{Kuznetsov:2006}, in that case, the contribution from the ground Landau level did not dominate due to the large electron virtuality, and contributions from other levels were of the same order. Ignoring this fact led the authors~\cite{Elizalde:2002,Elizalde:2004} to incorrect results.  Another example of this kind was an attempt to reanalyze the probability of the neutrino decay $\nu \to e^- W^+$ in an external magnetic field in the limit of ultra-high neutrino energies, calculated via the imaginary part of the one-loop amplitude of the transition $\nu \to e^- \, W^+ \to \nu$.
Initially, the result was obtained in Ref.~\cite{Erdas:2003}. Later, the calculation was repeated in ~\cite{Bhattacharya:2009} where authors insisted on another result. The third independent calculation \cite{Kuznetsov:2010_PLB} confirmed the result of Ref.~\cite{Erdas:2003}. The most likely cause of the error in Ref.~\cite{Bhattacharya:2009} was the use of only linear terms in the expansion of the $W$-boson propagator, whereas the quadratic terms were essential as well.

Landau-levels position-space representation is unique in the sense that it provides an explicit dependence of the propagator on the space-time coordinates, at the same time splitting the expression into individual terms according to the energy-related quantum numbers (Landau levels). This allows for the simultaneous treatment of the problem from both space-time and energetic perspectives.

This paper is structured as follows. In Chapter 2, in addition to the results of our previous work \cite{Iablokov:2020ypp}, we rederive the position-space Landau-levels representation of the charged scalar particle propagator in a constant magnetic field using three different methods, namely, the original Fock-Schwinger approach \cite{Schwinger:1951,Itzykson_1980}, the modified Fock-Schwinger approach \cite{MFS_2018,Iablokov:2020upc}, and the canonical quantization approach. Overall, this chapter serves as a brief review of methods for obtaining different representations of charged particle propagators in a constant magnetic field. These methods demonstrate a high degree of consistency and provide additional intermediate representations of the propagators. Based on the results of Chapter 2, in Chapters 3 and 4, we obtain previously unpublished Landau-levels position-space representations for the propagators of charged particles with spin, i.e., a fermion and a massive vector boson. In Chapter 5, we discuss the properties of the obtained expressions using the example of a scalar particle propagator.

Throughout calculations, we adhere to the "mostly minus" metric convention $g_{\mu\nu} = (+,-,-,-)$. 
The constant magnetic field $B$ is directed along the $z$-axis. This leads to the decomposition of the spacetime vectors into parallel ($\parallel$) and perpendicular ($\perp$) components
\begin{equation}
\centering
p_\perp^\mu = (0, p_1, p_2, 0) \, , \qquad p_\parallel^\mu = (p_0, 0, 0, p_3) \, ,
\label{eq:vect_prp_prl}
\end{equation}
which belong to the Euclidean \{1, 2\}-subspace and the Minkowski \{0, 3\}-subspace, 
correspondingly. Then, for the arbitrary four-vectors $p_\mu$, $q_\mu$ one has
\begin{equation}
(pq)_\perp = p_1 q_1 + p_2 q_2 \, , \qquad 
(pq)_\parallel = p_0 q_0 - p_3 q_3 \, ,
\label{eq:prod_prp_prl}
\end{equation}
with the full scalar product written as
\begin{equation}
(pq)=(pq)_\parallel - (pq)_\perp \, .
\end{equation}


\section{Charged scalar particle propagator}

\subsection{Proper-time representation}

In this section, we briefly outline the derivation of the proper-time position-space representation of a charged scalar particle propagator in a constant magnetic field using the original Fock-Schwinger (FS) approach. 

In the FS approach, we are to solve the following propagator equation:
\begin{equation}
\label{eq_FS_010}
H(\partial_X, X) \, G(X, X') = \delta^{(4)} (X-X') \, .
\end{equation}
The FS method consists in representing the unknown function $G(X,X')$ as an integral
\begin{equation}
\label{eq_FS_020}
G(X, X') = (-\I) \int\limits_{-\infty}^{0} \D \tau \,U(X,X';\tau) \, ,
\end{equation}
where $U(X,X';\tau)$ satisfies a Schr{\"o}dinger-type equation
\begin{equation}
\label{eq_FS_050}
\I \, \partial_\tau U(X, X';\tau) = H(\partial_X, X) \, U(X,X';\tau) \, 
\end{equation}
with the following boundary conditions:
\begin{equation}
\label{eq_FS_080}
U(X, X';0) = \delta^{(4)} (X-X') \, , \, \, U(X, X';-\infty) = 0 \, .
\end{equation}
The solution then reads:
\begin{equation}
\label{eq_FS_090}
U(X, X';\tau) = \E^{ -\I \, \tau H(\partial_X, X) + \varepsilon \tau} \delta^{(4)} (X-X') \, .
\end{equation}
The action of the exponential operator could be evaluated directly (as will be discussed later), however, the original FS method follows a different strategy. The relevant details could be found in \cite{Schwinger:1951,Itzykson_1980}.
Here, we provide just the final expression for the proper-time position-space representation of the charged scalar particle propagator in a constant magnetic field:
\begin{eqnarray}
\label{eq_FS_430}
G(X, X') &=& -\frac{\beta}{(4\pi)^2} \E^{\I\Phi(X,X') } \int\limits_{-\infty}^{0} \D \tau \, \frac{1}{\tau\sin(\beta\tau)} \quad \quad \quad 
\\
\nonumber
&\times& \exp{\I \bigg[ \frac{Z^2_\shortparallel}{4\tau} - \frac{\beta Z^2_\perp}{4\tan(\beta\tau)} + (m^2 - \I\varepsilon) \tau \bigg] } \, ,
\end{eqnarray}
where $Z^\mu = X^\mu - X'^\mu$ and 
\begin{eqnarray}
\label{eq_FS_phase}
\Phi(X,X') = - eQ \! \int^{X}_{X'} \! \D\xi^\mu \! \left [ A_\mu(\xi) \! + \! \frac{1}{2} F_{\mu\nu} (\xi - X')^\nu \right] \! .
\end{eqnarray}


\subsection{\label{sec:level2}Landau-levels representation}
In this section, we switch from the proper-time representation (\ref{eq_FS_430}) to the sum over Landau levels.

One can notice that the integral in (\ref{eq_FS_430}) resembles the well-known identity (see, e.g., \cite{GR}) for the modified Bessel functions of the second kind:
\begin{eqnarray}
\label{eq_Knu}
K_\nu(z) = \frac{1}{2} \left(\frac{z}{2} \right)^\nu \int^\infty_0 \frac{dt}{t^{\nu + 1}} \E^{-t-\frac{z^2}{4t}} \, ,
\end{eqnarray}
which is valid for $\Re \, z^2 > 0$ and $|\arg \, z| < \pi/4$.

To use (\ref{eq_Knu}), we transform (\ref{eq_FS_430}) according to the following recipe. First, a change of integration variable is performed:
\begin{eqnarray}
\label{eq_tau_to_t}
t \equiv -\I (m^2 - \I \varepsilon) \tau \, .
\end{eqnarray}
This effectively rotates the integration contour by the angle of $\approx -\pi / 2 - \varepsilon/m^2$ and leads to the following formula:
\begin{eqnarray}
\label{eq_G_t_1}
G(X,X') &=& \frac{\I\beta}{(4\pi)^2} \int^0_{(i+\varepsilon)\infty} \frac{dt}{t}  \E^{-\frac{(-Z^2_\shortparallel)(m^2-\I\varepsilon)}{4t} - t} 
\\
\nonumber
&\times& \!\! \bigg( \left[\sinh\left(\beta t / m^2 \right) \right]^{-1} \E^{-\frac{\beta Z^2_\perp}{4} \cotanh\left( \beta t / m^2 \right)} \bigg) \, .
\end{eqnarray}
The expression in brackets can be written as:
\begin{eqnarray}
\left[\sinh\left(\beta t / m^2 \right) \right]^{-1} \E^{-\frac{\beta Z^2_\perp}{4} \cotanh\left( \beta t / m^2 \right)} = \quad \quad \quad \quad \quad
\\
\nonumber
= 2 \frac{\E^{-b}}{1-c} \E^{a\frac{c+1}{c-1}} = 2 \frac{\E^{-b}\E^{-a}}{1-c} \E^{-\frac{2ac}{1-c}} \, ,
\end{eqnarray}
where $a=\beta Z^2_\perp / 4$, $b = t\beta / m^2$ and $c=\E^{-2b}$.

Transition to the sum over Landau levels is performed via the following formula for the generating function of Laguerre polynomials $L_n$, which is valid for $|c| < 1$:
\begin{eqnarray}
\label{eq_laguerre_gen_func}
\frac{1}{1-c} \E^{-\frac{2ac}{1-c}} = \sum_{n=0}^\infty L_n(2a) \, c^n \, .
\end{eqnarray}
We additionally change the contour of integration from $\left( (\I + \varepsilon)\infty , 0\right)$ to $(0, \infty)$, where the condition $c = e^{-2b}=e^{-2t\beta / m^2} < 1$ is satisfied.

After the straightforward rearrangements in the exponential factors in (\ref{eq_G_t_1}) and the following substitutions
\begin{eqnarray}
\label{eq_t_to_s}
\frac{t}{m^2} = \frac{s}{M_n^2} \, , \quad \quad M_n^2 = m^2 + (2n+1)\beta  \, ,
\end{eqnarray}
we use (\ref{eq_Knu}) and obtain for $Z^2_\shortparallel < 0$:
\begin{eqnarray}
\label{eq_G_K_L}
G(X,X') &=& -\frac{\I\beta}{4\pi^2}\E^{\I\Phi}\E^{-\beta Z^2_\perp / 4} 
\\
\nonumber
&\times& \sum^\infty_{n=0} K_0\left(M_n \sqrt{-Z^2_\shortparallel + \I\varepsilon}\right)\, L_n \left(\frac{\beta Z^2_\perp}{2}\right) \, . 
\end{eqnarray}
For the case $Z^2_\shortparallel > 0$, we transform $K_0$ using a standard relation for Bessel functions:
\begin{eqnarray}
\label{eq_knu_h2nu}
K_\nu(z) = -\frac{\I\pi}{2} \, \E^{-\I\pi\nu/2} H_{\nu}^{(2)}\bigg( z \E^{-\I \pi/2} \bigg) \, ,
\end{eqnarray}
where $H_\nu^{(2)}$ are the Hankel functions of the second kind.
The final expression for the position-space Landau-levels representation of the propagator reads:
\begin{eqnarray}
\label{eq_prop_scalar_final}
G(X, X') = \frac{-\I\beta}{4\pi^2} \, \E^{\I \Phi}  
\sum_{n=0}^{\infty} \, \Phi_n \, \bigg[ K_0 - \frac{\I\pi}{2} \, H_0^{(2)}  \bigg] \, ,
\end{eqnarray}
where the notations were used: 
\begin{eqnarray}
\label{eq_prop_scalar_final_notations}
\Phi_n &=& L_n  \E^{-\beta Z^2_\perp / 4} \, , \, \, L_n = L_n \bigg(\frac{\beta Z^2_\perp}{2}\bigg) \, ,
\\
K_0 &=& K_0 \bigg( M_n \sqrt{-Z_\parallel^2 + \I \epsilon} \bigg) \, \theta(-Z_\parallel^2) \, ,
\\
H_0^{(2)} &=& H_0^{(2)} \bigg( M_n \sqrt{Z_\parallel^2 - \I \epsilon} \bigg) \, \theta(Z_\parallel^2) \, .
\end{eqnarray}
\\
The light-cone behaviour of $K_0$ and $H^{(2)}_0$ is discussed, e.g., in \cite{Zhang:2008jy}.
\\


\subsection{Canonical quantization approach}
An alternative way to obtain an analytic expression for the propagator is to consider a time-ordered ``sum over the solutions'' of the corresponding wave-equation:
\begin{equation}
\label{eq_weq}
H(\partial_X,X) \psi(X) = 0 \, .
\end{equation}
The $H$ operator can be written as:
\begin{eqnarray}
\label{eq_weq_h}
H(\partial_X,X) &=& \Pi^\mu \Pi_\mu -m^2 
\\
\nonumber
&=& (\I\partial)^2_\shortparallel + \beta \left[ d^2_\eta - \eta^2 \right] - m^2 \, ,
\end{eqnarray}
where $\eta = \sqrt{\beta}\left( x - Q \frac{p_y}{\beta} \right)$ and $\beta = eB$.
It gives the following set of solutions:
\begin{equation}
\label{eq_weq_solutions}
\psi_n^{(\pm)}(X) = \frac{\beta^{1/4}}{\sqrt{2 p_0 L_y L_z}} \E^{\mp\I p_0 t + \I p_y y + \I p_z z} V_n(\eta) \, ,
\end{equation}
where 
\begin{equation}
\label{eq_p0}
p_0 = \sqrt{p^2_z + m^2 + (2n+1)\beta} 
\end{equation}
and $L_y/L_z$ are the volume normalization factors.
The functions $V_n$ are the quantum harmonic oscillator (QHO) eigenvectors 
\begin{equation}
\label{eq_Vn}
V_n(\eta) = \frac{1}{\sqrt{2^n n! \sqrt{\pi}}} \E^{-\eta^2/2} H_n(\eta) \, ,
\end{equation}
satisfying
\begin{equation}
\label{eq_Vn_eigen}
\left[ d^2_\eta - \eta^2 \right] V_n(\eta) = -(2n+1) V_n(\eta) \, ,
\end{equation}
with $H_n$ being the Hermite polynomials.

Performing standard calculations in the canonical quantization scheme, we obtain the propagator
\begin{equation}
\label{eq_prop_0}
G(X,X') = (-\I)\bra{0} T \{ \psi(X) \psi^*(X') \} \ket{0}
\end{equation}
as the ``sum over the solutions'':
\begin{eqnarray}
\label{eq_G_py}
G(X,X') &=& \sqrt{\beta} \sum^\infty_{n=0} \int \frac{d^2p_\shortparallel dp_y}{(2\pi)^3} 
\\
\nonumber
&\times&\frac{\E^{-\I(p(X-X'))_\shortparallel + \I p_y (y-y')}}{p^2_\shortparallel - M_n^2 +\I \varepsilon} V_n(\eta) V_n(\eta') \, ,
\end{eqnarray}
where $\eta' = \sqrt{\beta}\left( x' - Q \frac{p_y}{\beta} \right)$.
This form of the propagator, however, is not symmetric with respect to 
$x,y$ coordinates and, therefore, does not reflect the internal symmetry of the problem.
To symmetrize (\ref{eq_G_py}), we should perform the $p_y$-integration:
\begin{equation}
\label{eq_inn}
\begin{gathered}
I_{n,n'} = \int \D p_y e^{\I p_y(y-y')} V_n(\eta) V_{n'} (\eta') \, .
\end{gathered}
\end{equation}
First, we make a change of the integration variable:
\begin{equation}
\label{eq_7.2}
\begin{gathered}
u = -Q \frac{p_y}{\sqrt{\beta}} + \frac{\sqrt{\beta}}{2} \left[ (x+x') + \I Q (y-y') \right] \, .
\end{gathered}
\end{equation}
This leads to:
\begin{equation}
\label{eq_7.3}
\begin{gathered}
I_{n,n'} = \frac{\E^{\I \Phi(X,X')}}{\sqrt{2^{n+n'} n! n'! \, \pi}} \sqrt{\beta} \, \E^{-\frac{\beta}{4} (X-X')^2_\perp} \tilde{I}_{n,n'} \, ,
\end{gathered}
\end{equation}
where
\begin{eqnarray}
\label{eq_phase_xy}
\Phi(X,X') &=& \frac{Q\beta}{2} (x+x')(y-y') \, ,
\\
\tilde{I}_{n,n'} &=& \int_{-\infty}^\infty \D u \, \E^{-u^2} H_n (u+a) H_{n'} (u+b) \, ,
\end{eqnarray}
with the following substitutions:
\begin{eqnarray}
\nonumber 
a &=& \frac{\sqrt{\beta}}{2} \left[ (x-x') - \I Q (y-y') \right] \, ,
\\
b &=& - \frac{\sqrt{\beta}}{2} \left[ (x-x') + \I Q (y-y') \right] \, .
\end{eqnarray}
The phase (\ref{eq_phase_xy}) can be shown to be in agreement with the one in Eq.~(\ref{eq_FS_phase}),
see e.g. Ref.~\cite{KM_Book_2013}.
Second, according to Ref.~\cite{GR}, the $\tilde{I}_{n,n'}$ integral evaluates to:
\begin{eqnarray}
\label{eq_inn_tilde}
\nonumber
\tilde{I}_{n,n'} &=& 2^{n'} \sqrt{\pi} \, n! \, b^{n'-n} L^{(n'-n)}_{n}(-2ab) 
\\
&=& 2^{n'} \sqrt{\pi} \, n! \, b^{n'-n} L^{(n'-n)}_{n}\left(\frac{\beta}{2}Z^2_\perp \right) \, 
\end{eqnarray}
for $n \leq n'$.
Finally, we obtain the symmetrized representation of the propagator:
\begin{eqnarray}
\label{eq_prop_symm}
\nonumber
G(X, X') &=& \frac{\beta}{2 \pi} \E^{\I \Phi} \sum_{n=0}^\infty \, L_n \bigg(\frac{\beta Z^2_\perp}{2}\bigg) \, \E^{-\beta Z^2_\perp / 4} \,
\\
&\times& \int \frac{\D^2 p_{\,\shortparallel}}{(2\pi)^2} \frac{\E^{{- \I \left( pZ \right)}_{\parallel}}}{p^2_\shortparallel - M_n^2 + \I \varepsilon} \, .
\end{eqnarray}
We observe that, aside from the overall non-invariant phase factor $\E^{\I\Phi}$, the summation terms in (\ref{eq_prop_symm}) decompose into two factors. 
The first one depends only on the $x,y$-coordinates and is invariant with respect to rotations in the $x,y$-plane which is perpendicular to the direction of the magnetic field. The second factor (the Fourier integral) 
\begin{eqnarray}
J_{\parallel} = \int \frac{\D^2 p_{\,\shortparallel}}{(2\pi)^2} \frac{\E^{{- \I \left( pZ \right)}_{\parallel}}}{p^2_\parallel - M_n^2 + \I \varepsilon}
\end{eqnarray}
depends only on the $t,z$-coordinates and allows for further simplification.
In the case $Z^2_\shortparallel < 0$, it evaluates to:
\begin{eqnarray}
\label{eq_jk0}
J_{\parallel} = \frac{-\I}{2\pi } K_0 \bigg( M_n \sqrt{-Z_\parallel^2 + \I \epsilon} \bigg)  \, .
\end{eqnarray}
For $Z^2_\shortparallel > 0$, we once again apply (\ref{eq_knu_h2nu}), which gives the same final answer (\ref{eq_prop_scalar_final}) for the propagator.


\subsection{\label{sec:level2}Modified Fock-Schwinger method}

Yet another approach for finding propagators of charged particles in external electromagnetic fields, the modified Fock-Schwinger (MFS) method \cite{MFS_2018,Iablokov:2020upc}, consists in the direct evaluation of (\ref{eq_FS_090}). First, one should choose an appropriate representation of the $\delta$-function. For the considered problem, the following decomposition is particularly convenient:  
\begin{equation}
\label{delta_repr}
\delta^{(4)} (X-X') = \sqrt{\beta} \sum_{n=0}^\infty \int \frac{\D ^3 p_{\,\shortparallel, y}}{(2\pi)^3} \,
\E^{{-\I \left( pZ \right)}_{\shortparallel, y}} V_n (\eta) V_n (\eta ') \, .
\\
\end{equation}
Next, using (\ref{eq_FS_090}), (\ref{eq_weq_h}) and  (\ref{eq_Vn_eigen}), we write (\ref{eq_FS_020}) as:
\begin{eqnarray}
\label{eq_result_exp_h}
G(X, X') &=& - \I \sqrt{\beta} \sum_{n=0}^\infty \int \frac{\D ^3 p_{\,\shortparallel, y}}{(2\pi)^3} \, \E^{{-\I \left( pZ \right)}_{\shortparallel, y}}
\\
\nonumber
&& \int_{-\infty}^{0} \D \tau \,  \E^{-\I \tau \left( p^2_\parallel - M_n^2 + \I \varepsilon \right)}  V_n V_n ' \, ,
\\
\nonumber
\end{eqnarray}
which after integration over $\tau$ gives us (\ref{eq_G_py}). The rest of calculations are identical to (\ref{eq_inn}) -- (\ref{eq_jk0}).

In the case of a scalar particle, the MFS method gives no obvious advantage as compared to the calculations performed in the canonical quantization scheme. However, for particles with spin, it significantly simplifies the notation and provides additional representations of the propagator. It should also be noted that the effectiveness of the MFS method was demonstrated in a different physical scenario, namely, the method was applied by other authors to calculate the fermion propagator in a rotating environment \cite{ayala2021fermion}. 
We also believe that this approach could appear to be useful for the calculations of particle propagators in various physical environments. For example, in a self-dual field, very simple expressions for the scalar and fermion propagators exist. While the self-dual electromagnetic field (defined by the condition ${\vec E} = i {\vec H}$) could not be a non-zero physical field, its analysis could be useful as a preliminary step to more complicated non-abelian fields, e.g. the gluon field in QCD, see \cite{dubovikov1981analytical}. 

In the following chapters, we focus our attention exclusively on the calculations using the MFS method.


\section{\label{sec:level1} Charged fermion propagator}

The $H(\partial_X, X)$ operator in Eq.~(\ref{eq_FS_010}) for a charged fermion in a constant magnetic field reads:
\begin{eqnarray}
\label{eq_FS_H_fermion}
H(\partial_X, X) = \Pi_\mu \gamma^\mu - m \, .
\end{eqnarray}
Squaring the operator by the following substitution
\begin{eqnarray}
\label{eq_h_sqr}
G(X,X') = \left( \Pi_\mu \gamma^\mu + m \right) S(X,X') 
\end{eqnarray}
leads to the equation from which the new unknown function $S(X,X')$ could be determined:
\begin{eqnarray}
\label{eq_FS_3}
H(\partial_X, X) \, S(X, X') = \delta^{(4)} (X-X') \, ,
\end{eqnarray}
with $H$ redefined as:
\begin{eqnarray}
\label{eq_FS_3}
H(\partial_X, X) = \!\left[(\I\partial)^2_\shortparallel\!+\!\beta\!\left( d^2_\eta\!-\!\eta^2 \right)\!-\!m^2 \right]\!I\!+\!Q\beta \Sigma_3\, . 
\end{eqnarray}
Given that $\Sigma_3 = \I \gamma_1 \gamma_2 = \diag(+1,-1,+1,-1)$  commutes with the unit matrix $I$, we are able to separate the exponential operator in (\ref{eq_FS_090}) into two exponents:
\begin{eqnarray}
\label{eq_23}
\nonumber
S(X,X') &=& (-\I)\sqrt{\beta} \sum_{n=0}^\infty \int \frac{\D^2 p_{\shortparallel} \D p_y}{(2\pi)^3}\int_{-\infty}^0 
\D \tau 
\\
\nonumber
&\times& \E^{-\I \tau Q\beta \Sigma_3} \E^{-\I \tau\left[  p_{\shortparallel}^2 - \beta(2n+1) - m^2 + \I \ee \right]}
\\
&\times& \E^{{-\I \left( p(X-X') \right)}_{\shortparallel, y}} V_n(\eta) V_n(\eta ') \, .
\end{eqnarray}
%
Next, we evaluate $\exp (-\I\tau Q \beta \Sigma_3)$ and combine it with the second exponent, followed by the shift of the summation index where appropriate. This allows us to perform the proper-time integration and obtain the following expression:
\begin{eqnarray}
\label{eq_prop_f_symm}
\nonumber
S(X,X') &=& \sqrt{\beta} \,  \sum_{n=0}^\infty \int \frac{\D^2 p_{\shortparallel}\, \D p_y}{(2\pi)^3} \,
\frac{\E^{{-\I \left( p(X-X') \right)}_{\shortparallel, y}}}{p_{\shortparallel}^2 - M_n^2 + \I \ee} \, 
\\
&\times& \bigg[ V_{n-1} V_{n-1}^{'} \Pi^{(Q)}_{n-1} + V_{n} V_{n}^{'} \Pi^{(Q)}_{n} \bigg] \, ,
\end{eqnarray}
where 
\begin{eqnarray}
\label{eq_prop_f_projections}
\nonumber
\Pi^{(Q)}_{n-1} &=& \frac{1-Q}{2} \Pi_{+} + \frac{1+Q}{2} \Pi_{-} \, , 
\\
\nonumber
\Pi^{(Q)}_{n} &=& \frac{1+Q}{2} \Pi_{+} + \frac{1-Q}{2} \Pi_{-} \, , 
\\
\nonumber
\Pi_{+} &=& \diag(1,0,1,0) \, , 
\\
\nonumber
\Pi_{-} &=& \diag(0,1,0,1) \, ,
\end{eqnarray}
and $M_n^2 = m^2 + 2\beta n$.
Using intermediate expressions within the original FS approach \cite{Schwinger:1951,Itzykson_1980}, one can derive a useful relation:
\begin{eqnarray}
\label{eq_pi_to_fmunu}
&&\bigg(\I \partial_\mu - eQA_\mu \bigg)\left[ \E^{\I\Phi} f \right] = 
\\
\nonumber
&& \quad \quad \quad \quad \quad=\E^{\I\Phi} \left(\I \partial_\mu + \frac{Q\beta}{2} \varphi_{\mu\nu} (X-X')^\nu \right) f  \, . 
\end{eqnarray}
Following the same techniques as in the case of a scalar particle and applying (\ref{eq_pi_to_fmunu}) and (\ref{eq_h_sqr}), we obtain the final result:
\begin{eqnarray}
\label{eq_prop_fermion_almost_final}
\nonumber
G(X, X') &=& \frac{-\I\beta}{4\pi^2} \, \E^{\I \Phi} 
\bigg[ \left( \I \partial_\mu  + \frac{Q\beta}{2} \varphi_{\mu\nu} Z^\nu \right) \gamma^\mu + m \bigg] 
\\
\nonumber
&\times& \sum_{n=0}^{\infty} \E^{-\beta Z^2_\perp / 4} 
\bigg[ \, L_{n-1} \Pi^{(Q)}_{n-1} + L_{n} \Pi^{(Q)}_{n} \bigg] \, 
\\
&\times& \bigg( K_0  - \frac{\I\pi}{2} \, H_0^{(2)}  \bigg) \, . \quad \quad
\end{eqnarray}
Here, $\varphi^{\mu\nu} = F^{\mu\nu} / B$. We leave the representation (\ref{eq_prop_fermion_almost_final}) as is, not performing the evaluation of derivatives. In real calculations involving propagators, these derivatives are usually integrated by parts. 


\section{\label{sec:level1} Charged massive vector boson propagator}

Lastly, let us consider the charged vector boson. The corresponding propagator equation is given by:
\begin{eqnarray}
\label{eq_w_west}
&& H^\mu_{\,\,\, \nu} G^\nu_{\,\, \rho}(X,X') = \delta^\mu_{\,\,\, \rho}  \delta^4(X-X') \, ,
\\
&& H^\mu_{\,\,\, \nu} = \left( H_0 \right)^\mu_{\,\,\, \nu} + \left( H_F \right)^\mu_{\,\,\, \nu} + \left( H_\xi \right)^\mu_{\,\,\, \nu}  \, ,
\nonumber
\end{eqnarray}
where
\begin{eqnarray}
\label{eq_H_0_west}
\left( H_0 \right)^\mu_{\,\,\, \nu} \, &=& \left( \Pi \Pi - m^2 \right) \delta^\mu_{\,\,\, \nu} \, ,
\\
\label{eq_H_F_west}
\left(H_F \right)^\mu_{\,\,\, \nu} \, &=& - 2ieQF^\mu_{\,\,\, \nu} \, ,
\\
\label{eq_H_xi_west}
\left(H_\xi \right)^\mu_{\,\,\, \nu} \, &=& \left( \frac{1}{\xi} - 1  \right) \Pi^\mu \Pi_\nu \, .
\end{eqnarray}

We proceed with calculations exactly as in Ref.~\cite{Iablokov:2020upc}. Briefly, we note that $[H_0+H_F,H_\xi] = [H_0,H_F] = 0$. This allows for a step-by-step separation of the exponential operator $\E^{-\I \tau \left( H_0 + H_F + H_\xi \right)}$:
\begin{eqnarray}
\label{eq_7e}
\nonumber
\E^{-\I \tau \left( H_0 + H_F + H_\xi \right)} &=& \E^{-\I \tau H_\xi} \, \E^{-\I \tau \left( H_0 + H_F \right)} = \quad \quad \quad
\\
&=& \E^{-\I \tau H_\xi} \, \E^{-\I \tau H_F} \, \E^{-\I \tau H_0} \, .
\end{eqnarray}
Expanding the exponents 
\begin{eqnarray}
\label{eq_exp_H_F}
&&\left[\E^{-\I \tau H_F}\right]^\mu_{\,\,\, \nu} = \left[\E^{-2 Q \beta \tau \varphi}\right]^\mu_{\,\,\, \nu} = 
\quad \quad \quad
\\
\nonumber
&& \quad = \delta^\mu_{\parallel  \nu}\!+\!\frac{\E^{\I 2\beta \tau}}{2}\left( \delta^\mu_{\perp  \nu}\!+\!\I Q\varphi^\mu_{\,\,\, \nu} \right)\!+\!\frac{\E^{-\I 2\beta \tau}}{2}\left( \delta^\mu_{\perp  \nu}\!-\!\I Q\varphi^\mu_{\,\,\, \nu} \right) 
\end{eqnarray}
and
\begin{eqnarray}
\label{eq_7i}
 \left[\E^{-\I \tau H_\xi}\right]^\mu_{\,\,\, \nu} &=& \E^{-\I \tau \left( \frac{1}{\xi} - 1 \right) \Pi^\mu \Pi_\nu } = 
\\
\nonumber
 &=& \delta^\mu_{\,\,\, \nu} + \Pi^\mu \frac{ \E^{-\I \tau \left( \frac{1}{\xi} - 1 \right) \Pi \Pi  } - 1 }{\Pi\Pi} \Pi_\nu \, ,
\end{eqnarray}
we shift the summation index (as for the fermion case) and evaluate the corresponding $\tau$-integral. Performing the transformations (\ref{eq_inn}) -- (\ref{eq_inn_tilde}), we obtain the symmetrized representation of the propagator:
\begin{eqnarray}
\label{eq_prop_symm_w}
\nonumber
G^\mu_{\,\,\, \nu}(X, X') &=& \frac{\beta}{2 \pi} \E^{\I \Phi} \sum_{n=-1}^\infty \int \frac{\D^2 p_{\shortparallel}}{(2\pi)^2} \frac{\E^{{- \I \left( pZ \right)}_{\parallel}} \, \E^{-\beta Z^2_\perp / 4}}{p^2_\parallel - M^2_n + \I \varepsilon}
\\
&\times& \left( d^\mu_{\,\,\, \nu} + \frac{\xi - 1}{p^2_\parallel - \tilde{M}^2_n + \I \varepsilon} f^\mu_{\,\,\, \nu} \right) \, ,
\end{eqnarray}
\begin{eqnarray}
\label{eq_prop_w_d}
d^\mu_{\,\,\, \nu}  &=& \delta^\mu_{\parallel \nu} L_n + \frac{1}{2} \delta^\mu_{\perp \nu} \bigg ( L_{n+1} + L_{n-1} \bigg) \, ,
\\
\nonumber
&& \quad \quad \quad - \frac{\I Q}{2} \varphi^\mu_{\,\,\, \nu} \bigg( L_{n+1} - L_{n-1} \bigg) \, ,
\\
\label{eq_prop_w_f}
f^\mu_{\,\,\, \nu}  &=& \bigg[ p^\mu_\parallel p_{\parallel \nu} + \frac{\beta Q}{2} \bigg( p^\mu_\parallel 
(Z \varphi)_\nu + (\varphi Z)^\mu p_{\parallel \nu} \bigg) 
\\
\nonumber
&-& \bigg( (2n+1)\beta -\frac{\beta^2}{4}Z^2_\perp \bigg) \delta^\mu_{\perp \nu} + \frac{\I Q \beta}{2} \, \varphi^\mu_{\,\,\, \nu}  \bigg] L_n \, 
\\
\nonumber
&+& \frac{\I \beta}{2} \bigg[ \bigg( p^\mu_\parallel Z_{\perp \nu} + Z^\mu_\perp p_{\parallel \nu} \bigg) - \I \delta^\mu_{\perp \nu} - \frac{Q \beta}{2} Z^2_\perp \varphi^\mu_{\,\,\, \nu} \bigg] 
\\
\nonumber
&\times& \bigg( L^{(1)}_n + L^{(1)}_{n-1} \bigg) - \beta^2 (\varphi Z)^\mu (Z \varphi)_\nu L^{(2)}_{n-1} \, .
\end{eqnarray}
Here, $M^2_n = m^2 + (2n+1)\beta$, $\tilde{M}^2_n = \xi m^2 + (2n+1)\beta$ and all the Laguerre polynomials $L_n^{(m)}$ have $\beta Z_\perp^2 / 2$ as their arguments: $L_n^{(m)} = L_n^{(m)} (\beta Z_\perp^2 / 2)$.

Next, we have to consider both $J_\parallel$ and a similar integral
\begin{eqnarray}
\label{eq_J_parallel_xi}
\tilde{J}_{\parallel} &=& \int \frac{\D^2 p_{\,\shortparallel}}{(2\pi)^2} \frac{\E^{{- \I \left( pZ \right)}_{\parallel}}}
{\left[ p^2_\parallel - M_n^2 + \I \varepsilon \right]\left[ p^2_\parallel - \tilde{M}_n^{2} + \I \varepsilon \right]} 
\\
\nonumber
&=&
\frac{\I}
{2\pi} \frac{1}{m^2 (\xi - 1)} \, \bigg[ K_0  - \frac{\I\pi}{2} \, H_0^{(2)} - \tilde{K}_0  + \frac{\I\pi}{2} \, \tilde{H}_0^{(2)} \bigg] \, ,
\end{eqnarray}
where $\tilde{K}_0 = K_0 \bigg( \tilde{M}_n \sqrt{-Z_\parallel^2 + \I \epsilon} \bigg) \, \theta(-Z_\parallel^2)$ and $\tilde{H}_0^{(2)} = H_0^{(2)} \bigg( \tilde{M}_n \sqrt{Z_\parallel^2 - \I \epsilon} \bigg) \, \theta(Z_\parallel^2)$. For the ground-state level, one can notice that large field values lead to the vacuum instability \cite{Nielsen:1978rm}.

Replacing $p_\parallel$ by $\I \partial_\parallel$ in (\ref{eq_prop_w_f}), we finally obtain the position-space representation of the charged massive vector boson propagator as an expansion over Landau levels:
\begin{eqnarray}
G^\mu_{\,\nu}(X,\!X')\!=\!\frac{-\I\beta}{4 \pi^2}\!\E^{\I \Phi}\!\E^{-\beta Z^2_\perp / 4}\!\sum_{n=-1}^\infty\!\bigg\{ P^\mu_{n\, \nu}\!-\!\frac{1}{m^2}\!Q^\mu_{n\, \nu}\!\bigg\} \, , \, \, \,
\end{eqnarray}
\begin{eqnarray}
P^\mu_{n\, \nu} &=& \bigg[ \delta^\mu_{\parallel \nu} L_n + \frac{1}{2} \delta^\mu_{\perp \nu} \left ( L_{n+1} + L_{n-1} \right) 
\\
\nonumber
&& \quad - \frac{\I Q}{2} \varphi^\mu_{\,\,\, \nu} \left( L_{n+1} - L_{n-1} \right) \bigg] \bigg[ K_0  - \frac{\I\pi}{2} \, H_0^{(2)} \bigg] \, ,
\\
\nonumber
Q^\mu_{n\, \nu} &=& \bigg[
\bigg\{ -\partial^\mu_\parallel \partial_{\parallel \nu} + \frac{\I \beta Q}{2} \left( \partial^\mu_\parallel 
(Z \varphi)_\nu + (\varphi Z)^\mu \partial_{\parallel \nu} \right) 
\\
\nonumber
&& \quad  - \left( (2n+1)\beta -\frac{\beta^2}{4}Z^2_\perp \right) \delta^\mu_{\perp \nu} + \frac{\I Q \beta}{2} \, \varphi^\mu_{\,\,\, \nu}  \bigg\} L_n
\\
\nonumber
&& \quad  - \frac{\beta}{2}\!\bigg\{\!\left(\!\partial^\mu_\parallel Z_{\perp \nu}\!+\!Z^\mu_\perp \partial_{\parallel \nu}\!\right)\!-\!\delta^\mu_{\perp \nu}\!+\!\frac{\I Q \beta}{2}\!Z^2_\perp \varphi^\mu_{\,\,\, \nu}\!\bigg\} 
\\
\nonumber
&& \quad \times \bigg( L^{(1)}_n + L^{(1)}_{n-1} \bigg) 
- \beta^2 (\varphi Z)^\mu (Z \varphi)_\nu L^{(2)}_{n-1} \bigg]
\\
&\times& \bigg[ K_0  - \frac{\I\pi}{2} \, H_0^{(2)} - \tilde{K}_0  + \frac{\I\pi}{2} \, \tilde{H}_0^{(2)} \bigg]  \, .
\end{eqnarray}
%


\section{\label{sec:level1} Discussion}


In this chapter, we discuss the position-space Landau-levels representation of the obtained propagators. To simplify the analysis, we restrict our attention to the scalar particle propagator. 

First of all, we notice that each expansion term in (\ref{eq_prop_scalar_final}) is a product of two factors that correspond to the propagation in parallel and perpendicular directions with respect to the field. The Euclidean $x,y$-plane and Minkowsky $t,z$-plane are independent, yet, they are connected through the Landau level $n$ in the following sense. Each function 
\begin{eqnarray}
\label{eq_x}
\Phi_n(\alpha) = L_n(\alpha) \E^{-\alpha/2} \quad \quad \alpha = \beta Z^2_\perp / 2
\end{eqnarray}
describes perpendicular propagation, with its graph consisting of two regions, oscillatory and monotonic (see Fig. 1).
In the monotonic region, the damping exponential dominates, thus, making the respective contribution of the $n$-th Landau level to the total propagation amplitude negligible. 
\begin{figure}
\includegraphics[scale=0.50]{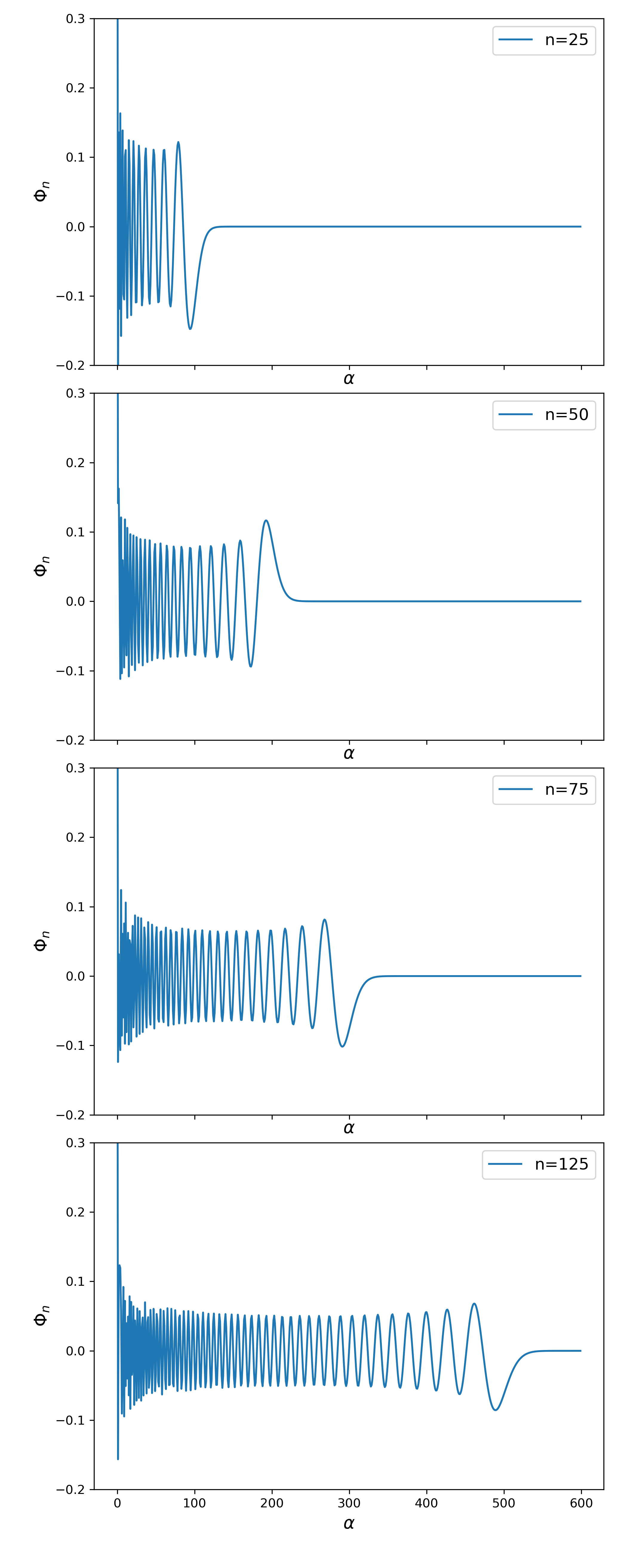}
\caption{Oscillatory behaviour of the functions $\Phi_n$ for various values of the Landau level $\left(n=25,50,75,125\right)$. The oscillatory region is bounded by $\alpha \approx 4n$ .}
\end{figure}
The oscillatory region has the following bounds \cite{temme1995uniform}:
\begin{eqnarray}
\label{eq_oscillatory_zone_bounds}
\alpha_{min} \approx 0 \, ,
\alpha_{max} \approx 4 n \, ,
\end{eqnarray}
and $\Phi_n$ inside this region is approximated by:
\begin{eqnarray}
\label{eq_oscillatory_zone_approx}
\Phi_n(\alpha) \approx \sqrt{\frac{2}{\pi}} \frac{\sin \mu (\alpha)}{ {\big[(\alpha_{max} - \alpha)(\alpha - \alpha_{min}) \big]}^{1/4} } \, ,
\end{eqnarray}
where $\mu(\alpha)$ is some function which exact form is not relevant for the present discussion (for details, see Ref.~ \cite{berry2008exact}).
This observation shares some similarity to the hydrogen atom problem where distant orbits are described by large principal quantum numbers.

From the formulas above, we first conclude that in order to describe perpendicular propagation up to 
the radial distance $R$ (i.e., up to $\alpha=\beta R^2/2$) we need to consider at least $n_{min}$ expansion terms, where
\begin{eqnarray}
\label{eq_nmin}
n_{min} \approx \alpha / 4 \approx \beta R^2/8 \, .
\end{eqnarray}
Second, the amplitude of oscillations of $\Phi_n(\alpha)$ in the oscillatory region (taken, e.g., at $\alpha_{max} / 2 = 2n$) depends on $n$ according to the following relation:
\begin{eqnarray}
\label{eq_x}
\Phi_n \sim \frac{1}{\sqrt{n}} \, .
\end{eqnarray}
This, together with (\ref{eq_nmin}), implies that the scale of the first expansion term with a non-vanishing value of $\Phi_n$ for a given perpendicular propagation distance $R$ depends on the radial distance as follows:
\begin{eqnarray}
\label{eq_x}
\Phi_n \sim \frac{1}{\sqrt{\beta} R} \, .
\end{eqnarray}
%

%
%

It turns out that this is not the only way the radial distance $R$ affects the total propagation amplitude. The Landau level $n$ also enters the $M_n$ parameter, which $K_0$ and $H^{(2)}_0$ depend on. As we established in (\ref{eq_nmin}), large values of $R$ imply large values of $n_{min}$, which in turn results in large values of $\zeta = \sqrt{m^2 + (2n_{min}+1)\beta}\sqrt{| Z^2_\parallel |}$.
Using the known asymptotic relations for Bessel functions \cite{GR,PBM}
\begin{eqnarray}
\label{eq_x}
K_0 (\zeta) &\approx& \sqrt{\frac{\pi}{2\zeta}} \E^{-\zeta} \, ,
\\
H^{(2)}_0 (\zeta) &\approx& \sqrt{\frac{2}{\pi\zeta}} \E^{-\I(\zeta - \pi/4)} \, ,
\end{eqnarray}
we observe that the perpendicular propagation distance $R$ additionally controls the overall scale of the total propagation amplitude through parallel propagation factors 
according to the following asymptotic relations:
\begin{eqnarray}
\label{eq_x}
K_0 &\sim& \E^{-\beta R} / \sqrt{\beta R} \, ,
\\
H^{(2)}_0 &\sim& 1/\sqrt{\beta R} \, .
\end{eqnarray}
The obtained expressions might serve as an analytical tool for making decisions regarding the truncation of the series in various physical scenarios. 

\section{\label{sec:level1} Conclusion}

We obtained the position-space Landau-levels representations for the propagators of charged particles (scalar, fermion, and massive vector boson) in a constant magnetic field. For the scalar case, three discussed methods, i.e., the original Fock-Schwinger (FS) approach, the modified Fock-Schwinger (MFS) approach, and the canonical quantization approach, showed consistent results. Another popular strategy, namely, performing a Fourier transform of a known expression for a proper-time momentum-space representation, is also applicable to this problem. However, due to the need for a simultaneous evaluation of both momentum-space and proper-time integrals, it presents a challenging (however, feasible) task, especially for the vector boson case. In this paper, we demonstrated how one can omit such lengthy calculations by either focusing on the proper-time integral or the Fourier integrals. In particular, when applying the FS method (section 2.2), we started from a position-space representation and only needed to perform the transition from the proper-time integral to the Landau-levels series. At the same time, in the MFS method (section 2.4), the early (and simple) evaluation of the proper-time integral left us with three Fourier transforms to be performed. In both cases, the overall computational complexity was significantly reduced.

The obtained Landau-levels expansions of the propagators have a simple structure, with each summation term represented as a product of two factors. The first one depends only on the coordinates of the plane perpendicular to the direction of the magnetic field. It is also rotationally invariant, thus, emphasizes the underlying symmetry of the problem. Being a product of a Laguerre polynomial and a damping exponential, this factor localizes the propagation in the $x,y$-plane. The second factor corresponds to the propagation in a 1+1--dimensional space-time, and is invariant with respect to Lorentz transformations in this subspace. Similar to the case of a free field, it contains both a time-like oscillatory term $H_0^{(2)}$ and a space-like damping term $K_0$. 

These representations are unique in the sense that they allow for the simultaneous study of the propagator from both space-time and energetic perspectives. This is encoded in the number of the Landau level, which is deeply connected with the radial propagation distance in the perpendicular plane. 

We expect further use of the obtained position-space Landau-levels representations, e.g., in the study of loop processes, in finite-spacetime calculations, and in various astrophysical problems.

\begin{acknowledgements}
We are grateful to A.\,Ya.~Parkhomenko and D.\,A.~Rumyantsev for useful remarks. 

The reported study was funded by RFBR, project number \mbox{19-32-90137}. 
\end{acknowledgements}

\bibliographystyle{utphys}       
\bibliography{coord-epjc}   

\end{document}